\documentclass[journal=jacsat,manuscript=article]{achemso}

\usepackage{chemformula} 
\usepackage[T1]{fontenc} 
\usepackage{xcolor}
\usepackage{setspace}
\usepackage{graphicx}
\graphicspath{ {./figures/} }
\usepackage{subcaption}
\usepackage{amsmath}
\usepackage{siunitx}
\usepackage{layouts}
\usepackage{color,soul}
\usepackage{booktabs}

\usepackage[colorinlistoftodos]{todonotes}

\author{C\'edric~Lemieux-Leduc}
\author{Mahmoud~R.~M.~Atalla}
\author{Simone~Assali}
\author{Nicolas~Rotaru}
\author{Julien~Brodeur}
\author{Stéphane~Kéna-Cohen}
\author{ Oussama~Moutanabbir}
\email{oussama.moutanabbir@polymtl.ca}
\author{ Yves-Alain~Peter}

\affiliation{Department of Engineering Physics, \'Ecole Polytechnique de Montr\'eal, Montr\'eal, C.P. 6079, Succ. Centre-Ville, Montr\'eal, Qu\'ebec, Canada H3C 3A7}

\title{Waveguide-Coupled Mid-Infrared GeSn Membrane Photodetectors on Silicon-on-Insulator}

\begin{document}

\maketitle

\begin{abstract}

Silicon photonics has thrived in telecommunications over recent decades, and its extension to the mid-infrared range has the potential to unlock valuable opportunities for sensing, imaging, and free-space communications. With this perspective, germanium-tin (GeSn) alloy has been extensively investigated as a silicon-compatible semiconductor with bandgap tunability that covers this entire spectral range. Indeed, a variety of GeSn-based high-performance optoelectronic devices have been demonstrated, confirming the potential of this system for mid-infrared applications. However, the integration of these devices onto silicon photonic platforms remains underexplored. Herein, we demonstrate the fabrication and integration, through transfer-printing, of strain-relaxed GeSn membranes onto silicon-on-insulator waveguides to create integrated detectors operating up to 3.1~$\mu$m at room temperature. Two different designs of waveguide structures are evaluated to study the coupling efficiency between the passive structures and the active membrane detector. A responsivity reaching 0.36~A/W at an operation wavelength of 2.33~$\mu$m is measured under a bias of 1~V. Moreover, the fabrication resulted in multiple working devices exhibiting similar performance using a single transfer printing step, demonstrating the scalability of the proposed approach. 

\end{abstract}



\section*{Introduction}

The field of silicon photonics has attracted substantial interest driven by the ability to implement photonic integrated circuits (PICs), supporting a variety of applications ranging from optical data communication to chemical and biomedical sensing~\cite{thomson2016roadmap, shekhar2024roadmapping}. These circuits are fabricated using wafer-scale complementary metal-oxide-semiconductor (CMOS) processing techniques, facilitating low-cost and high-volume production of PICs. Predominantly, these chips are based on the silicon-on-insulator (SOI) platform, where the high refractive index contrast between silicon and silicon dioxide supports the development of very compact and high-density passive waveguide circuits. Initially, advances in this research field were largely driven by applications in telecommunications, particularly around the 1.3~$\mu$m and 1.5~$\mu$m wavelengths. However, there is an interest in developing the operational capacity of PICs towards longer wavelengths in the mid-infrared range (2--8~$\mu$m), with a particular interest around the 2~$\mu$m band to expand the capacity for intrachip communications~\cite{soref2015enabling, hu2017silicon, zou2018midinfrared}. This shift is also motivated by the potential for new applications, such as integrated sensing systems for environmental monitoring and industrial process control, where the strong absorption peaks of various molecules are located within this spectral range~\cite{sieger2016toward, yu2018siliconchipbased, wang2018widely}.

Moreover, the mid-infrared region is suitable for integrated quantum photonics in silicon due to reduced two-photon absorption (TPA) and improved Kerr effect~\cite{rosenfeld2020midinfrared}. Despite these advancements, achieving a fully integrated system requires active elements such as light emitters and photodetectors on the same chip along with passive components. These active components must meet specific performance criteria, including sufficient output power for light emitters and high responsivity combined with a broad spectral range for photodetectors. However, silicon and germanium, which are typically found in CMOS processes, have an indirect bandgap and with energies of 1.12~eV and 0.67~eV respectively, making them unsuitable for photodetection in the mid-infrared range. Si avalanche photodetectors that rely on subbandgap transitions from lattice defects showed extended operation up to 1.96~$\mu$m with responsivity of 0.3~A/W and a bit rate of 20~Gb/s, though the applied voltage bias above 15~V for avalanche mode is relatively high~\cite{ackert2015highspeed}.

In parallel, remarkable progress has been made in the development of III--V compound semiconductors for active devices operating in the mid-infrared range, particularly with materials such as indium gallium arsenide (InGaAs) with an increased amount of indium (extended-InGaAs) and gallium antimonide (GaSb), which have demonstrated cutoff wavelengths exceeding 2.3~$\mu$m~\cite{yin2007review, lei2008lasers}. These materials are typically integrated into passive circuits through hybrid or heterogeneous integration methods, as direct growth of these compounds on silicon poses significant challenges. Successful integration of lasers and photodetectors with silicon-based waveguides operating in this wavelength range has been achieved through the heterogeneous integration of these compounds on silicon~\cite{Annual2010, gassenq2012study, wang2016IIIVonsilicon, spott2017heterogeneous, wang2019widely}. As a cost-effective alternative, germanium-tin (GeSn), a group-IV isovalent alloy, has been subject of extensitve research  for mid-infrared optoelectronics exploiting its tunable bandgap energy ~\cite{moutanabbir2021monolithic, reboud2024advances}. Indeed, recent GeSn photodetectors have demonstrated impressive high-speed performance, with reported bandwidths up to 50~GHz and cutoff wavelengths above 2.3~$\mu$m~\cite{cui2024highspeed, atalla2024extended}. However, GeSn integration with silicon photonics remains underexplored \cite{wang2022monolithic} and mostly limited to numerical investigations~\cite{li2020design, ghosh2022design}. GeSn waveguide photodetector devices have been demonstrated and characterized, but as monolithic devices through edge coupling~\cite{tsai2021gesn, wang2021highspeed}. However, dislocations at the interface of monolithic devices lead to high optical losses, which can directly affect the device performance. Circumventing these challenges can be, in principle, achieved through the heterogeneous integration with the SOI platform. GeSn layer transfer using bonding has already been demonstrated, resulting in a GeSn-on-insulator (GeSnOI) platform or GeSn on a Germanium-on-insulator (GeOI) platform~\cite{xu2019integrating, Joo2021, Jung22}. Alternately, GeSn can be released selectively to create strain-relaxed membranes or microdisks, which can then be transferred onto different substrates~\cite{atalla2021allgroup, kim2022, chen2022transferable, lemieuxleduc2025transferprinted}. Photodetectors based on these transferred structures exhibit increased detection cutoff wavelengths as a result of relaxation of the compressive strain. Transferrable membranes are potentially suitable for the heterogeneous integration of the alloy with silicon photonics, with transfer printing emerging as an effective approach to the integration of active materials into passive PICs where direct growth can be impractical~\cite{carlson2012transfer, yoon2015heterogeneously, degroote2016transferprinting, zhang2019iiivonsi, haq2020microtransferprinted}. This process combines the benefits of flip-chip integration and wafer bonding, enabling the co-integration of various active materials onto the same PIC.

Herein, using transfer printing, we demonstrate waveguide-integrated photodetectors based on Ge$_{0.89}$Sn$_{0.11}$ membranes, with a detection cutoff of 3.1~$\mu$m at room temperature, placed atop SOI waveguides. Contacts are then evaporated to achieve metal-semiconductor-metal (MSM) photodetectors, which operate in the mid-infrared and at room temperature. Before proceeding with device fabrication, modeling based on finite-difference time-domain (FDTD) calculations is performed to identify and optimize critical design parameters influencing the coupling between the waveguide and the membrane. Several devices have been fabricated, where two designs for the passive SOI structures are tested to evaluate the impact of coupling between the waveguide and the GeSn material. The first design is a 2~$\mu$m-wide waveguide, which solely relies on evanescent coupling to enable the transfer of light from the waveguide to the GeSn detector. The second design is based on a tapered waveguide with a diffraction grating, which would include a contribution of grating-assisted coupling in addition to evanescent coupling. The characterization of these devices at a wavelength of 2.33~$\mu$m reveals a responsivity of up to 0.36~A/W under a bias of 1~V. Multiple fabricated devices exhibit similar properties, which demonstrate the viability of the approach based on transfer printing to enable the integration of GeSn onto silicon photonic platforms.

\section*{Results and discussion}

\noindent {\bf Design and modeling} 

\begin{figure}[htbp]
    \centering
    \includegraphics[width=\textwidth]{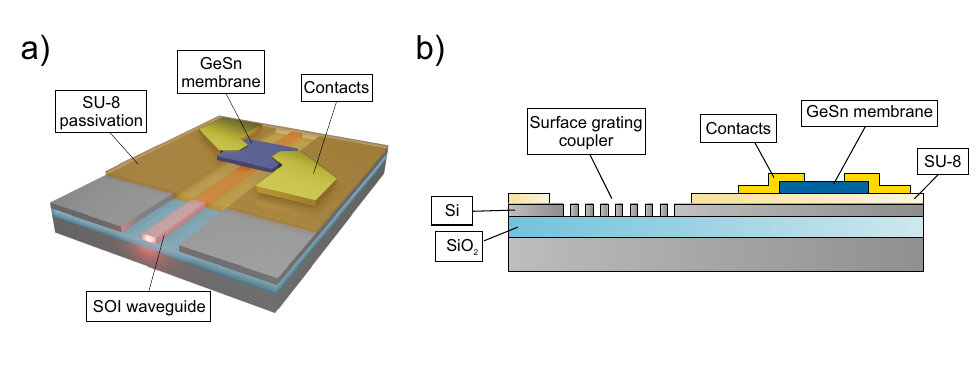}
    \caption{Device schematic of a waveguide-integrated GeSn photodetector. a) Illustration of the concept. b) Cross-section of the designed device. A GeSn membrane is transfer-printed on top of a waveguide using a thin SU-8 layer to enhance adhesion and passivate the circuit.}
    \label{fig:concept}
\end{figure}

Figure~\ref{fig:concept}(a) illustrates the proposed concept of a transfer-printed GeSn membrane onto an SOI waveguide, forming an integrated MSM mid-infrared detector. The device relies on the evanescent coupling between the waveguide and the GeSn membrane. Light is coupled into the waveguide via a surface grating coupler (SGC). An SU-8 layer is spin-coated on the passive chip to enhance the adhesion of the membrane to the chip during transfer printing and provide electrical insulation for the metal contacts. Once the GeSn membranes are transferred onto the passive chip, a Ti/Au bilayer is evaporated to form contacts and complete the MSM devices. Figure~\ref{fig:concept}(b) illustrates the cross-section of the complete device, highlighting the different layers involved. Openings in SU-8 are introduced around the SGCs to maintain air as the top cladding, ensuring that the transmission properties remain independent of the SU-8 thickness. The GeSn membranes, which are approximately 500~nm-thick, have a content of 11~at.\%~Sn. These strain-relaxed membranes exhibit a photodetection wavelength cutoff of roughly 3.1~$\mu$m \cite{lemieuxleduc2025transferprinted}. 

\begin{figure}[htbp]
    \centering
    \includegraphics[width=\textwidth]{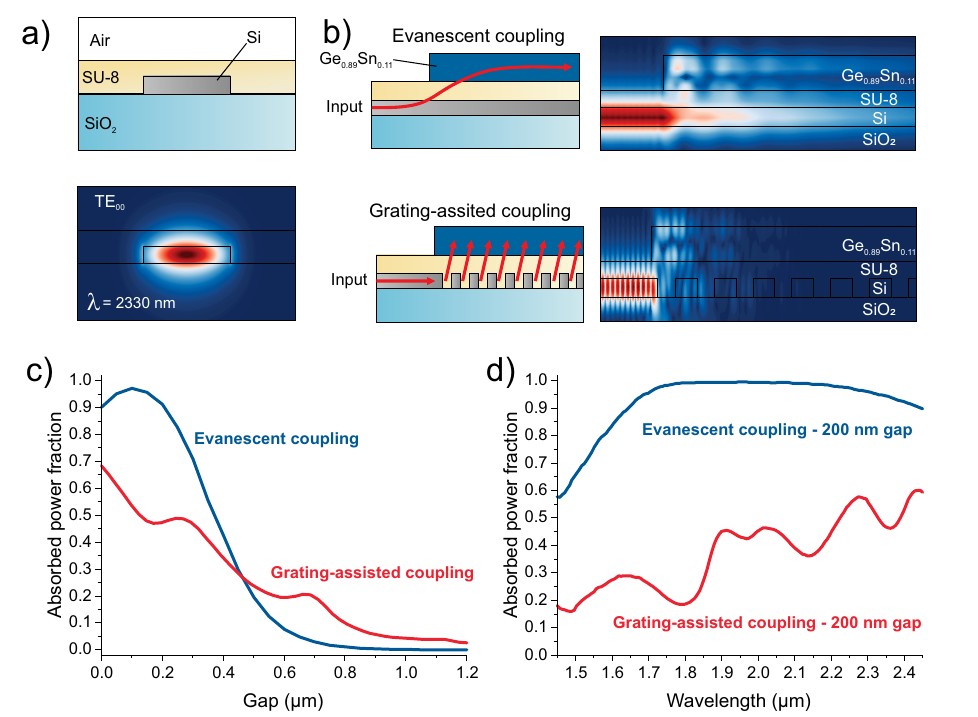}
    \caption{Modeling results of the evanescent and grating-assisted coupling between an SOI waveguide and a Ge$_{0.89}$Sn$_{0.11}$ membrane. (a) Cross-section of the waveguide (top) with the normalized distribution of $|E|$ of TE$_{00}$ at a wavelength of 2.33~$\mu$m (bottom). (b) Illustration of the principle of evanescent (top) and grating-assisted coupling (bottom) with the distribution of $|E|$ of the propagating mode in each case. Evolution of the absorbed power fraction for the two coupling mechanisms as a function of (c) the separation gap and (d) the wavelength.}
    \label{fig:simulations}
\end{figure}

To elucidate the interaction between the optical waveguide and the membrane and determine the range of optimal separation gaps, FDTD simulations were conducted. Figure~\ref{fig:simulations} provides an overview of the modeling results for an SOI waveguide interacting with a Ge$_{0.89}$Sn$_{0.11}$ membrane. The optical constants for SU-8 and Ge$_{0.89}$Sn$_{0.11}$ were obtained using spectroscopic ellipsometry (SE) up to 2.5~$\mu$m (see SI for details regading the measurements). For Ge$_{0.89}$Sn$_{0.11}$, the values were extracted from an as-grown sample. However, it is crucial to note that strain relaxation caused by the complete underetching of the membranes from the thin film alters the band structure, increasing the absorption cutoff wavelength from 2.8~$\mu$m to 3.1~$\mu$m, and causing an overall increase in absorption and a shift in the refractive index values. Figure~\ref{fig:simulations}(a) shows the cross-section of an SOI waveguide with a 200 nm~thick layer of SU-8 and the distribution of the electric field of the fundamental transverse electric (TE) mode TE$_{00}$ supported by the waveguide. The waveguides are designed with a width of 2~$\mu$m and are fully etched down to the buried oxide layer (BOX) to form strip waveguides with a 220~nm height and ensure operation with a TE-polarization. The effective refractive index $n_{eff}$ of this fundamental mode at a wavelength of 2.33~$\mu$m is 2.43392 and has an effective area of 0.66~$\mu$m$^2$. Although higher-order modes can propagate through the waveguide, as the geometry was not optimized for single-mode operation, these should not have a significant impact on the measurements carried out on the integrated device. The SGCs are designed to operate around an operation wavelength of 2.33~$\mu$m, and FDTD simulations were performed to determine the optimal design parameters. For injection at a \qty{10}{\degree} angle on an SOI platform with air as the top cladding, the optimization through modeling yields a grating period of 1.22~$\mu$m, a fill factor of 0.66, and an etch depth of 220~nm. The etch depth simplifies the fabrication as all the structures can be fully etched in a single step.

The performance of the device is intrinsically linked to its ability to efficiently transfer energy from the waveguide to the membrane detector. Therefore, design parameters must be carefully optimized to enhance the interaction between these two structures. Evanescent coupling, also known as directional coupling, occurs when two guiding media are close. The supported modes in each medium overlap through their respective evanescent tails, facilitating energy transfer from one mode to another. In this context, the waveguide is transparent with a refractive index of approximately 3.4, whereas the membrane has a refractive index of $\sim$4.3 and significant absorption at 2.33~$\mu$m. The SU-8 layer, positioned between the two guiding media, has a lower refractive index of approximately 1.58 and is transparent. The coupling gap is determined by its thickness, which can be adjusted by the spin coating speed and its solids' content. The efficiency of evanescent coupling decreases exponentially as the separation gap increases, since the overlap between the fields of the supported modes diminishes. Consequently, small thicknesses on the order of hundreds of nanometers are typically required to achieve efficient coupling. An alternative approach is using a diffraction grating to transfer light from the waveguide to the active medium, known as grating-assisted coupling. This type of structure is more tolerant to bigger gaps and relaxes the requirement for thin adhesive layers, but the efficiency is highly dependent on the quality of the grating and its performance. Figure~\ref{fig:simulations}(b) illustrates the two coupling mechanisms with a propagation profile computed using 2D FDTD modeling, showing the norm of the electric field $|E|$ and highlighting the interaction between the waveguide and a 20~$\mu$m-long membrane. The gap between the membrane and the waveguide structures is fixed at 200~nm. As the light travels from left to right and interacts with the membrane, its intensity decreases due to the absorption by the Ge$_{0.89}$Sn$_{0.11}$ membrane. In simulations involving grating-assisted coupling (GAC), the same design parameters as the SGCs are used for the diffraction grating. In the field distribution of evanescent coupling, oscillations are observed due to a small coupling length resulting from a mismatch between the propagation constants in each structure at this wavelength. Engineering the effective refractive indices can help achieve phase-matching, which would increase the coupling length and lead to full absorption of light over a smaller length within a given wavelength range. In the field distribution of the grating-assisted coupling, the propagating light appears to be absorbed over a shorter distance, but it is scattered at the angle of the diffraction grating, with backward reflections being present. These effects reduce the efficiency of the coupling scheme.

Figure~\ref{fig:simulations}(c) illustrates the fraction of the total power of the injected light absorbed as a function of the gap for a fixed wavelength of 2.33~$\mu$m, considering a 20~$\mu$m-long membrane for both mechanisms. For evanescent coupling, this fraction decreases rapidly as the gap increases, highlighting the sensitivity of this mechanism to small separations. In contrast, the grating-assisted coupling shows a steadier decrease in absorption efficiency with increasing gap thickness and can still absorb some power at gaps exceeding 500~nm. For both coupling methods, optimal thicknesses range from 100 to 300~nm. Figure~\ref{fig:simulations}(d) presents the fraction of power absorbed by the membrane as a function of wavelength for a fixed gap of 200~nm. The evanescent coupling scheme maintains its absorption efficiency over a broad wavelength range. This is because the decrease in Ge$_{0.89}$Sn$_{0.11}$ absorption is counterbalanced by the expansion of the waveguide mode towards longer wavelengths. Alternately, the efficiency of the grating-assisted coupling mechanisms is more variable but peaks around the design wavelength of 2.33~$\mu$m. Optimizing the design of the grating coupler could further enhance the performance of grating-assisted coupling~\cite{gassenq2012study}. Additionally, with a thickness of 200~nm, there may be a contribution from evanescent coupling to the overall absorption.

\noindent {\bf Growth of the GeSn layer and device fabrication}

\begin{figure}[htbp]
    \centering
    \includegraphics[width=\textwidth]{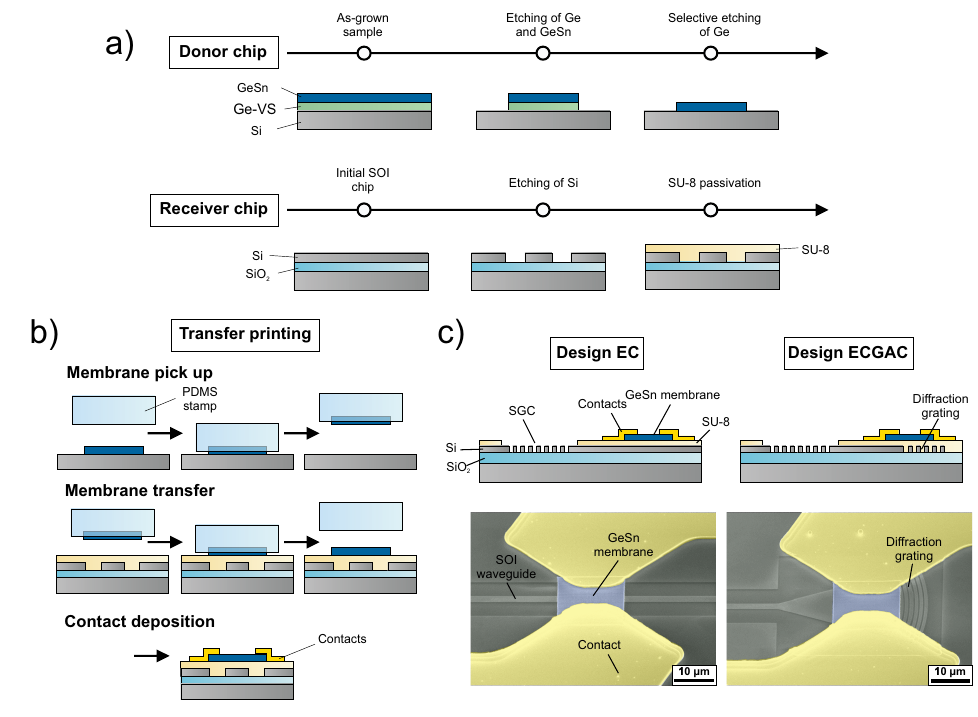}
    \caption{Fabricated chip. a) Fabrication steps of the donor and receiver chips. b) Transfer printing steps with the subsequent deposition of metal contacts on the transferred membranes. c) Sketches of the cross-section and SEM micrographs of the two waveguide structure designs studied, with the evanescent coupling design, EC (left) being a membrane atop a straight waveguide and grating-assisted coupling design, ECGAC (right) a tapered waveguide and a diffraction grating.}
    \label{fig:fab}
\end{figure}

Figure~\ref{fig:fab} provides an overview of the fabrication steps of the integrated membrane detectors with SEM micrographs of the results. The use of transfer printing requires a separate processing of the Ge$_{0.89}$Sn$_{0.11}$ membranes (donor chip) and the passive SOI chip containing the waveguides (receiver chip). Figure~\ref{fig:fab}(a) summarizes these steps for both chips. For the donor chip, a 530 nm thick GeSn layer is initially grown on a Si (100) substrate using chemical vapor deposition (CVD) on top of a 1~$\mu$m-thick Ge interlayer, also referred to as Ge virtual substrate (Ge-VS). X-ray diffraction reciprocal space mapping (XRD-RSM) indicates that the GeSn layer contains approximately 11~at.\%~Sn and has a biaxial compressive strain of about -0.6\%. An array of 20~$\mu$m~$\times$~20~$\mu$m released membranes is obtained by etching through the GeSn and Ge layers using a solution based on ceric ammonium nitrate and nitric acid. The Ge-VS is then selectively etched using reactive ion etching (RIE) with a carbon tetrafluoride (CF$_4$) plasma, resulting in free-standing Ge$_{0.89}$Sn$_{0.11}$ membranes. The characterization of these membranes has been reported previously~\cite{lemieuxleduc2025transferprinted}, showing an extension of the cutoff wavelength from 2.8~$\mu$m to 3.1~$\mu$m when compared to photodetectors based on the as-grown stack. Independently, the receiver chip is patterned on an SOI chip (B-doped, 8.5--\qty{11.5}{\ohm\cdot\centi\meter}) with a 220~nm device layer using electron beam lithography (EBL) with a 20~KV acceleration and using PMMA 495k A6 for patterning and masking. The layout consists of an array of 2~$\mu$m wide waveguides coupled to SGCs, with the pitch matching that of the membrane array to enable the fabrication of multiple devices at once. The etching and patterning of the silicon device layer are performed by RIE using SF$_6$ and C$_4$F$_8$ gases. The insertion loss of a single SGC is estimated to be around 12.9~dB at 2.33~$\mu$m. No stitching defects were observed in the waveguide structures, leading to the assumption that the propagation loss is negligible compared to the insertion loss of the SGCs. The chip is then passivated by spin coating a 160~nm-thick layer of SU-8, which is diluted in cyclopentanone to achieve the desired thickness. Openings in this layer around the SGCs are patterned using UV exposure to maintain an air cladding for coupling.

Ge$_{0.89}$Sn$_{0.11}$ membranes are then transferred onto the passive chip using a transfer printing setup (HQ Graphene) inside a glovebox with a nitrogen atmosphere. Figure~\ref{fig:fab}(b) illustrates the various steps involved in the transfer printing of the Ge$_{0.89}$Sn$_{0.11}$ membranes. A $\sim$1.5~mm~$\times$~1.5~mm polydimethylsiloxane (PDMS) stamp, controlled by a multiaxis stage, is used to pick up multiple Ge$_{0.89}$Sn$_{0.11}$ membranes by applying pressure on the donor chip. Once the membranes are picked up, the donor chip is replaced with the receiver chip underneath the PDMS stamp. The membranes are then transferred from the stamp to the receiver chip by applying pressure and then slowly retracting the stamp. This process ensures that the membranes detach from the PDMS stamp and adhere to the SU-8 adhesion layer on the receiver chip. After the transfer of the membranes is complete, the chip undergoes a thermal treatment at \qty{250}{\celsius} for 30 min under a nitrogen atmosphere to hardbake SU-8 and thermally anneal the Ge$_{0.89}$Sn$_{0.11}$ membranes. The hardbake of SU-8 helps maximize the density of crosslinks in the polymer matrix. The chip with the transferred membranes is then dipped in a mixture of HF:HCl:H$_2$O for 15~s to remove residual byproducts from the underetching step on the membrane surface, such as nonvolatile SnF$_x$ species. Subsequently, pairs of metal contacts with 20~nm of Ti and 120~nm of Au are evaporated onto the membranes using a lift-off technique. Prior to the evaporation of metal, the sample undergoes a 30-second O$_2$ plasma to enhance the adhesion of the metal on SU-8. This results in photoconductive Ge$_{0.89}$Sn$_{0.11}$ membranes coupled to waveguide structures. A gap between the contacts of approximately 7~$\mu$m is obtained.

The array of waveguided structures consists of two distinct designs of waveguides to accommodate the transferred membranes: one where a Ge$_{0.89}$Sn$_{0.11}$ membrane is aligned on top of a waveguide (evanescent coupling) and another where the membrane is placed on top of a grating coupler (grating-assisted coupling). The waveguide length between the SGC and the membrane is 125~$\mu$m. Figure~\ref{fig:fab}(c) presents sketches of the cross-section (top) and scanning electron microscopy (SEM) micrographs (bottom) of the two designs of the fabricated chips. The images demonstrate good adhesion of the membranes and metal contacts to the underlying structures. The evanescent coupling  design (left) features a membrane directly transferred onto a waveguide. The grating-assisted coupling design (right) consists of a tapered waveguide to increase the modal overlap with the Ge$_{0.89}$Sn$_{0.11}$ membrane and incorporates a diffraction grating to enhance responsivity through grating-assisted coupling. Due to the narrow gap provided by the SU-8 thickness, it is expected that there is a contribution of evanescent coupling combined with grating-assisted coupling in this design. Processing the devices as an array enabled the fabrication of multiple devices based on the two designs at once.

\noindent {\bf Device characterization}

\begin{figure}[htbp]
    \centering
    \includegraphics[width=\textwidth]{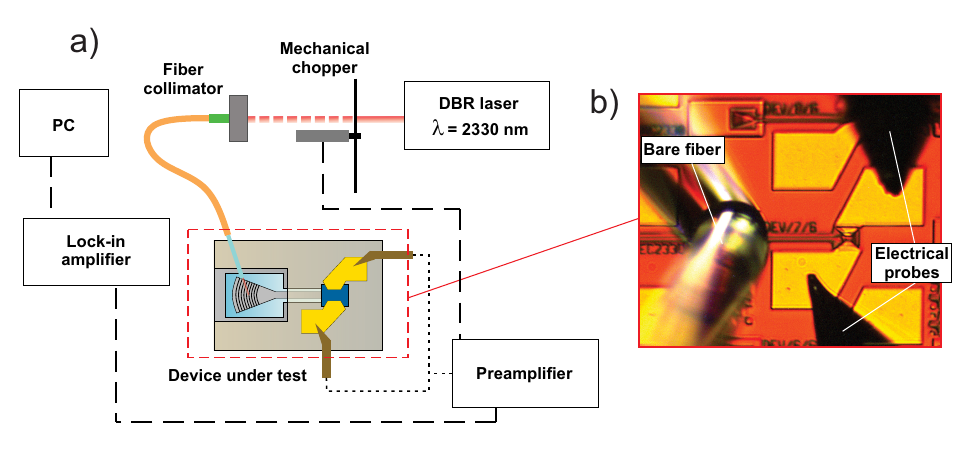}
    \caption{Characterization setup. (a) Schematic of the setup components to measure the responsivity while injecting light. (b) Microscope image of a coupled device with electrical probes.}
    \label{fig:setup}
\end{figure}

The devices are characterized using the setup illustrated in Fig.~\ref{fig:setup}. A multiaxis stage and a microscope are employed to precisely position an SMF-28 optical fiber above the SGC at the optimal injection angle. A custom distributed Bragg reflector (DBR) laser with a center wavelength of 2.33~$\mu$m and TE-polarization (custom laser from Nanoplus) is collimated using a FiberPort collimator (PAF2P-11E, Thorlabs) for injection into the optical fiber. Although the propagation loss at this wavelength is significant in an SMF-28 optical fiber, the optical power at the end of the bare fiber is set as reference and measured using an extended-InGaAs biased photodetector (DET10D2, Thorlabs). Electrical probes are placed on the metal pads to collect the photocurrent, which is measured using a lock-in technique. The light source is modulated at 500~Hz using a mechanical chopper, and the signal is analyzed with a lock-in amplifier (MFLI, Zurich Instruments). A current preamplifier (SR570, Stanford Research Systems) is used to amplify the signal and apply a bias voltage to the MSM device. This technique eliminates dark current and minimizes noise components in the measured signal. The setup was calibrated using the extended-InGaAs biased photodetector. Considering the insertion losses from the fiber to the waveguide via the SGC, which are approximately 12.9~dB, an estimated maximal optical power of 92~$\mu$W can be injected into the waveguide (after the SGC) before it starts interacting with the membrane. 

\begin{figure}[htbp]
    \centering
    \includegraphics[width=\textwidth]{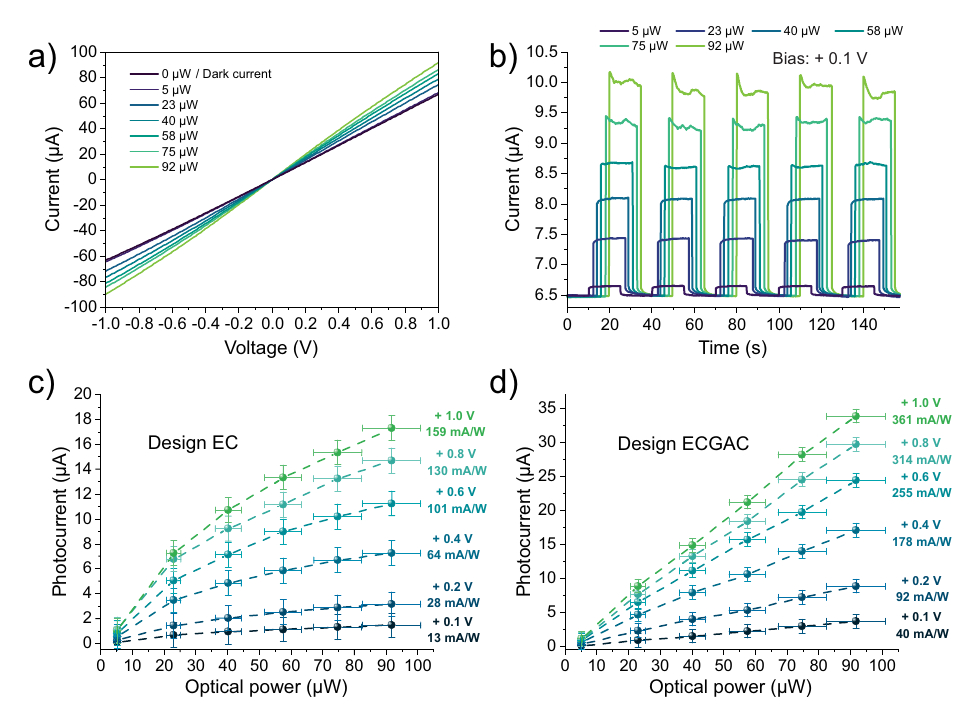}
    \caption{Characterization of the devices. (a) Current-voltage measurements on a device from -1~V to 1~V under illumination. (b) Transient photoresponse of a device under illumination at different optical powers. Photocurrent as a function of optical power in the waveguide for a device based on (c) Design EC and (d) Design ECGAC.}
    \label{fig:characterization}
\end{figure}

Figure~\ref{fig:characterization} summarizes the device characterization results. In Fig.~\ref{fig:characterization}(a), the current-voltage curves are shown with and without illumination on a device based on grating-assisted coupling design (Design ECGAC), using a semiconductor analyzer (Keithley 4200A-SCS) from -1~V to 1~V. The dark current is approximately 66~$\mu$A at 1~V with a linear behavior that indicates Ohmic-like contacts. This contrasts with the case where the contacts are also transferred rather than evaporated (see SI for details). Under illumination, the total current increases as a function of the optical power of the light being injected into the waveguide and the applied bias voltage, indicating that photocurrent is being generated and collected. Figure~\ref{fig:characterization}(b) presents the time-dependent response of a device under a 0.1~V bias voltage when exposed to different optical powers. The device responds relatively quickly, with a consistent total current overall, and returns to the original dark current value when illumination is removed. However, above a power of 58~$\mu$W, the total current appears to be unstable, possibly due to thermal effects caused by the high power density. 

Figures~\ref{fig:characterization}(c) and \ref{fig:characterization}(d) show the photocurrent as a function of optical power injected into the waveguide for applied biases from 0.1~V to 1~V for devices fabricated following the two designs. The bias was limited to 1~V to avoid degradation of the devices. Under evanescent coupling (Design EC), the photocurrent response exhibits a significant nonlinearity with increasing optical power, whereas Design ECGAC demonstrates a more linear response. This nonlinearity may originate from an electric field screening effect induced by carrier trapping, as the response tends toward a more linear behavior with increasing bias~\cite{mao2004fabrication}.
The Design ECGAC device achieved a responsivity of 361 mA/W at a 1 V bias, compared to 159 mA/W for the Design EC device. The error on the computed optical power inside the waveguide is estimated to be roughly 10\% of the computed value based on the instrumentation and the insertion loss variation in SGCs. The photocurrent error bar reflects the maximum oscillation amplitude observed at the highest injected optical powers, which is conservatively estimated as three times this amplitude and then rounded to 1~$\mu$A.

\begin{figure}[htbp]
    \centering
    \includegraphics[width=\textwidth]{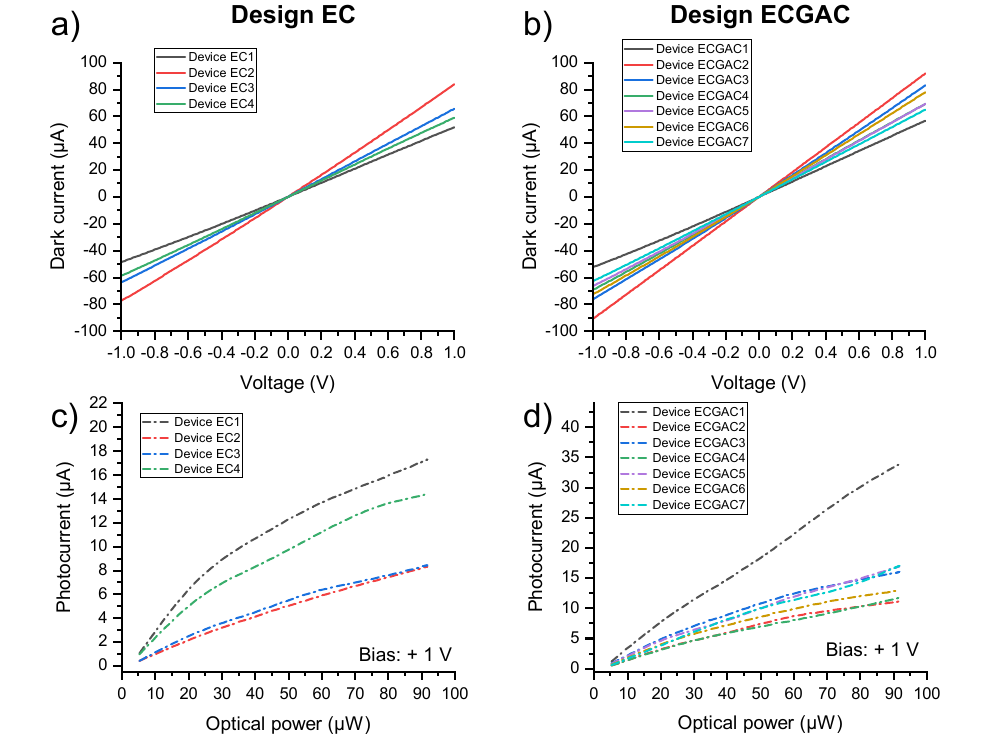}
    \caption{Overview of the characterization of the other devices. Dark current measurements of devices based on the (a) EC and (b) ECGAC designs. Photocurrent measurements at a bias of 1~V of devices based on the (c) EC and (d) ECGAC designs.}
    \label{fig:characterization2}
\end{figure}

To evaluate performance consistency, the rest of the fabricated devices were characterized. Figures~\ref{fig:characterization2}(a) and \ref{fig:characterization2}(b) show the dark current measured from -1~V to 1~V for Design EC and Design ECGAC, respectively. The devices exhibit linear behavior in each case, with dark currents ranging between 50 and 90~$\mu$A at a 1~V bias. The photocurrent as a function of optical power was also measured, with the results presented in Fig.~\ref{fig:characterization2}(c) and Fig.~\ref{fig:characterization2}(d) for each design. Despite minor variations in responsivity, all devices exhibit photocurrent within the same order of magnitude, with a responsivity lying mainly between 0.1 and 0.2~A/W while a maximal responsivity of 0.36 A/W is observed. A sublinear increase in the photocurrent is observable in devices based on either design. These variations may stem from differences in contact resistance, potentially caused by slight misalignment during membrane transfer, which can alter the effective contact area between the metal and the Ge$_{0.89}$Sn$_{0.11}$ membrane. Additional factors, such as the membrane's position relative to the waveguide or grating and variations in the SU-8 thickness, could also contribute to the discrepancy between simulation and experimental results. In fact, the simulations in Fig.~\ref{fig:simulations} predict a lower absorption efficiency when using a diffraction grating. To assess whether absorption efficiency differs between the two designs, 3D FDTD simulations were performed to analyze both instantaneous 2D and cumulative generation rates as a function of distance (see SI for details). However, the final cumulative generation rates showed only minimal differences. Evanescent coupling is likely the dominant mechanism in both cases, particularly in the Design ECGAC, where the membrane substantially overlaps with the tapered waveguide section.

Despite being waveguide-coupled, these detectors exhibit responsivity values comparable to most GeSn-based photodetectors~\cite{atalla2022highbandwidth, atalla2024extended}, with a maximum responsivity of 361~mA/W observed at a wavelength of 2.33~$\mu$m for GeSn. While p-i-n devices based on III--V exhibit an improved performance in terms of responsivity and dark current~\cite{hattasan2011heterogeneous, gassenq2012study, wang2016IIIVonsilicon}, these results highlight the potential of group-IV materials as a viable alternative for heterogeneously integrated photonic components operating in the mid-infrared region through the transfer printing of these structures.

\section*{Conclusion}

In this work, we demonstrated the integration of GeSn membranes onto SOI waveguides to create MSM detectors operating in the mid-infrared range. Using transfer-printed Ge$_{0.89}$Sn$_{0.11}$ strain-relaxed membranes, we achieved waveguide-coupled photodetectors operating at wavelengths up to 3.1~$\mu$m at room temperature. Different waveguide designs were simulated and tested to evaluate the coupling efficiency between the waveguide and the membrane. In contrast to simulations, experimental results showed that tapered waveguides with a diffraction grating yield higher responsivity as compared to direct coupling for 2~$\mu$m-wide waveguides. The recorded responsivity reaches 0.36~A/W at an operation wavelength of 2.33~$\mu$m under a bias of 1~V. The discrepancy between modeling and measurements is not fully understood and may perhaps originate from combined contributions of evanescent and grating-assisted coupling. The integration of mid-infrared photodetectors to PICs using GeSn membranes offers a cost-effective solution for portable spectroscopic systems and free-space communications. Although our approach involves transferring a large array of membranes simultaneously, an optimized transfer printing process that enables the picking and placing of individual membranes could help enhance the accuracy of this integration method.

\section{Supporting Information}

Details regarding ellipsometry measurement with the optical properties of Ge$_{0.89}$Sn$_{0.11}$ and SU-8; I--V comparison between evaporated and transfer-printed contacts; further simulation results for the generation rate in the final fabricated structures using 3D FDTD.

\section*{Acknowledgments}

The authors thank J. Bouchard for technical support with the CVD system. The authors also acknowledge support from NSERC Canada (Discovery and Alliance Grants), Canada Research Chairs, Canada Foundation for Innovation, Mitacs, PRIMA Québec, Defence Canada (Innovation for Defence Excellence and Security, IDEaS), the European Union’s Horizon Europe research and innovation programme under grant agreement No 101070700 (MIRAQLS), the US Army Research Office Grant No. W911NF-22-1-0277, and the Air Force Office of Scientific Research Grant No. FA9550-23-1-0763. C.L.-L. acknowledges financial support from NSERC (CGS D) and Fonds de recherche du Québec: Nature and Technologies (FRQNT, doctoral scholarship: https://doi.org/10.69777/272409). 

\bibliography{refs}

\providecommand{\latin}[1]{#1}
\makeatletter
\providecommand{\doi}
  {\begingroup\let\do\@makeother\dospecials
  \catcode`\{=1 \catcode`\}=2 \doi@aux}
\providecommand{\doi@aux}[1]{\endgroup\texttt{#1}}
\makeatother
\providecommand*\mcitethebibliography{\thebibliography}
\csname @ifundefined\endcsname{endmcitethebibliography}
  {\let\endmcitethebibliography\endthebibliography}{}
\begin{mcitethebibliography}{42}
\providecommand*\natexlab[1]{#1}
\providecommand*\mciteSetBstSublistMode[1]{}
\providecommand*\mciteSetBstMaxWidthForm[2]{}
\providecommand*\mciteBstWouldAddEndPuncttrue
  {\def\EndOfBibitem{\unskip.}}
\providecommand*\mciteBstWouldAddEndPunctfalse
  {\let\EndOfBibitem\relax}
\providecommand*\mciteSetBstMidEndSepPunct[3]{}
\providecommand*\mciteSetBstSublistLabelBeginEnd[3]{}
\providecommand*\EndOfBibitem{}
\mciteSetBstSublistMode{f}
\mciteSetBstMaxWidthForm{subitem}{(\alph{mcitesubitemcount})}
\mciteSetBstSublistLabelBeginEnd
  {\mcitemaxwidthsubitemform\space}
  {\relax}
  {\relax}

\bibitem[Thomson \latin{et~al.}(2016)Thomson, Zilkie, Bowers, Komljenovic,
  Reed, Vivien, Marris-Morini, Cassan, Virot, F{\'{e}}d{\'{e}}li, Hartmann,
  Schmid, Xu, Boeuf, O'Brien, Mashanovich, and Nedeljkovic]{thomson2016roadmap}
Thomson,~D. \latin{et~al.}  Roadmap on silicon photonics. \emph{Journal of
  Optics} \textbf{2016}, \emph{18}, 073003\relax
\mciteBstWouldAddEndPuncttrue
\mciteSetBstMidEndSepPunct{\mcitedefaultmidpunct}
{\mcitedefaultendpunct}{\mcitedefaultseppunct}\relax
\EndOfBibitem
\bibitem[Shekhar \latin{et~al.}(2024)Shekhar, Bogaerts, Chrostowski, Bowers,
  Hochberg, Soref, and Shastri]{shekhar2024roadmapping}
Shekhar,~S.; Bogaerts,~W.; Chrostowski,~L.; Bowers,~J.~E.; Hochberg,~M.;
  Soref,~R.; Shastri,~B.~J. Roadmapping the next generation of silicon
  photonics. \emph{Nature Communications} \textbf{2024}, \emph{15}, 751\relax
\mciteBstWouldAddEndPuncttrue
\mciteSetBstMidEndSepPunct{\mcitedefaultmidpunct}
{\mcitedefaultendpunct}{\mcitedefaultseppunct}\relax
\EndOfBibitem
\bibitem[Soref(2015)]{soref2015enabling}
Soref,~R. Enabling 2 $\mu$m communications. \emph{Nature Photonics}
  \textbf{2015}, \emph{9}, 358--359\relax
\mciteBstWouldAddEndPuncttrue
\mciteSetBstMidEndSepPunct{\mcitedefaultmidpunct}
{\mcitedefaultendpunct}{\mcitedefaultseppunct}\relax
\EndOfBibitem
\bibitem[Hu \latin{et~al.}(2017)Hu, Dong, Luo, Liow, Song, Lee, and
  Lo]{hu2017silicon}
Hu,~T.; Dong,~B.; Luo,~X.; Liow,~T.-Y.; Song,~J.; Lee,~C.; Lo,~G.-Q. Silicon
  photonic platforms for mid-infrared applications [Invited]. \emph{Photonics
  Research} \textbf{2017}, \emph{5}, 417\relax
\mciteBstWouldAddEndPuncttrue
\mciteSetBstMidEndSepPunct{\mcitedefaultmidpunct}
{\mcitedefaultendpunct}{\mcitedefaultseppunct}\relax
\EndOfBibitem
\bibitem[Zou \latin{et~al.}(2018)Zou, Chakravarty, Chung, Xu, and
  Chen]{zou2018midinfrared}
Zou,~Y.; Chakravarty,~S.; Chung,~C.-J.; Xu,~X.; Chen,~R.~T. Mid-infrared
  silicon photonic waveguides and devices [Invited]. \emph{Photonics Research}
  \textbf{2018}, \emph{6}, 254\relax
\mciteBstWouldAddEndPuncttrue
\mciteSetBstMidEndSepPunct{\mcitedefaultmidpunct}
{\mcitedefaultendpunct}{\mcitedefaultseppunct}\relax
\EndOfBibitem
\bibitem[Sieger and Mizaikoff(2016)Sieger, and Mizaikoff]{sieger2016toward}
Sieger,~M.; Mizaikoff,~B. Toward On-Chip Mid-Infrared Sensors. \emph{Analytical
  Chemistry} \textbf{2016}, \emph{88}, 5562--5573\relax
\mciteBstWouldAddEndPuncttrue
\mciteSetBstMidEndSepPunct{\mcitedefaultmidpunct}
{\mcitedefaultendpunct}{\mcitedefaultseppunct}\relax
\EndOfBibitem
\bibitem[Yu \latin{et~al.}(2018)Yu, Okawachi, Griffith, Picqu{\'{e}}, Lipson,
  and Gaeta]{yu2018siliconchipbased}
Yu,~M.; Okawachi,~Y.; Griffith,~A.~G.; Picqu{\'{e}},~N.; Lipson,~M.;
  Gaeta,~A.~L. Silicon-chip-based mid-infrared dual-comb spectroscopy.
  \emph{Nature Communications} \textbf{2018}, \emph{9}, 110803\relax
\mciteBstWouldAddEndPuncttrue
\mciteSetBstMidEndSepPunct{\mcitedefaultmidpunct}
{\mcitedefaultendpunct}{\mcitedefaultseppunct}\relax
\EndOfBibitem
\bibitem[Wang \latin{et~al.}(2018)Wang, Sprengel, Vasiliev, Boehm, Campenhout,
  Lepage, Verheyen, Baets, Amann, and Roelkens]{wang2018widely}
Wang,~R.; Sprengel,~S.; Vasiliev,~A.; Boehm,~G.; Campenhout,~J.~V.; Lepage,~G.;
  Verheyen,~P.; Baets,~R.; Amann,~M.-C.; Roelkens,~G. Widely tunable 2.3 $\mu$m
  {III}-V-on-silicon Vernier lasers for broadband spectroscopic sensing.
  \emph{Photonics Research} \textbf{2018}, \emph{6}, 858\relax
\mciteBstWouldAddEndPuncttrue
\mciteSetBstMidEndSepPunct{\mcitedefaultmidpunct}
{\mcitedefaultendpunct}{\mcitedefaultseppunct}\relax
\EndOfBibitem
\bibitem[Rosenfeld \latin{et~al.}(2020)Rosenfeld, Sulway, Sinclair, Anant,
  Thompson, Rarity, and Silverstone]{rosenfeld2020midinfrared}
Rosenfeld,~L.~M.; Sulway,~D.~A.; Sinclair,~G.~F.; Anant,~V.; Thompson,~M.~G.;
  Rarity,~J.~G.; Silverstone,~J.~W. Mid-infrared quantum optics in silicon.
  \emph{Optics Express} \textbf{2020}, \emph{28}, 37092\relax
\mciteBstWouldAddEndPuncttrue
\mciteSetBstMidEndSepPunct{\mcitedefaultmidpunct}
{\mcitedefaultendpunct}{\mcitedefaultseppunct}\relax
\EndOfBibitem
\bibitem[Ackert \latin{et~al.}(2015)Ackert, Thomson, Shen, Peacock, Jessop,
  Reed, Mashanovich, and Knights]{ackert2015highspeed}
Ackert,~J.~J.; Thomson,~D.~J.; Shen,~L.; Peacock,~A.~C.; Jessop,~P.~E.;
  Reed,~G.~T.; Mashanovich,~G.~Z.; Knights,~A.~P. High-speed detection at two
  micrometres with monolithic silicon photodiodes. \emph{Nature Photonics}
  \textbf{2015}, \emph{9}, 393--396\relax
\mciteBstWouldAddEndPuncttrue
\mciteSetBstMidEndSepPunct{\mcitedefaultmidpunct}
{\mcitedefaultendpunct}{\mcitedefaultseppunct}\relax
\EndOfBibitem
\bibitem[Yin and Tang(2007)Yin, and Tang]{yin2007review}
Yin,~Z.; Tang,~X. A review of energy bandgap engineering in III–V
  semiconductor alloys for mid-infrared laser applications. \emph{Solid-State
  Electronics} \textbf{2007}, \emph{51}, 6--15\relax
\mciteBstWouldAddEndPuncttrue
\mciteSetBstMidEndSepPunct{\mcitedefaultmidpunct}
{\mcitedefaultendpunct}{\mcitedefaultseppunct}\relax
\EndOfBibitem
\bibitem[Lei and Jagadish(2008)Lei, and Jagadish]{lei2008lasers}
Lei,~W.; Jagadish,~C. Lasers and photodetectors for mid-infrared $2-3$ $\mu$m
  applications. \emph{Journal of Applied Physics} \textbf{2008}, \emph{104},
  091101\relax
\mciteBstWouldAddEndPuncttrue
\mciteSetBstMidEndSepPunct{\mcitedefaultmidpunct}
{\mcitedefaultendpunct}{\mcitedefaultseppunct}\relax
\EndOfBibitem
\bibitem[Moutanabbir and Gösele(2010)Moutanabbir, and Gösele]{Annual2010}
Moutanabbir,~O.; Gösele,~U. Heterogeneous Integration of Compound
  Semiconductors. \emph{Annual Review of Materials Research} \textbf{2010},
  \emph{40}, 469--500\relax
\mciteBstWouldAddEndPuncttrue
\mciteSetBstMidEndSepPunct{\mcitedefaultmidpunct}
{\mcitedefaultendpunct}{\mcitedefaultseppunct}\relax
\EndOfBibitem
\bibitem[Gassenq \latin{et~al.}(2012)Gassenq, Hattasan, Cerutti, Rodriguez,
  Tourni{\'{e}}, and Roelkens]{gassenq2012study}
Gassenq,~A.; Hattasan,~N.; Cerutti,~L.; Rodriguez,~J.~B.; Tourni{\'{e}},~E.;
  Roelkens,~G. Study of evanescently-coupled and grating-assisted GaInAsSb
  photodiodes integrated on a silicon photonic chip. \emph{Optics Express}
  \textbf{2012}, \emph{20}, 11665\relax
\mciteBstWouldAddEndPuncttrue
\mciteSetBstMidEndSepPunct{\mcitedefaultmidpunct}
{\mcitedefaultendpunct}{\mcitedefaultseppunct}\relax
\EndOfBibitem
\bibitem[Wang \latin{et~al.}(2016)Wang, Muneeb, Sprengel, Boehm, Malik, Baets,
  Amann, and Roelkens]{wang2016IIIVonsilicon}
Wang,~R.; Muneeb,~M.; Sprengel,~S.; Boehm,~G.; Malik,~A.; Baets,~R.;
  Amann,~M.-C.; Roelkens,~G. {III}-V-on-silicon
  2-$\mathrm{\mu}$m-wavelength-range wavelength demultiplexers with
  heterogeneously integrated {InP}-based type-{II} photodetectors. \emph{Optics
  Express} \textbf{2016}, \emph{24}, 8480\relax
\mciteBstWouldAddEndPuncttrue
\mciteSetBstMidEndSepPunct{\mcitedefaultmidpunct}
{\mcitedefaultendpunct}{\mcitedefaultseppunct}\relax
\EndOfBibitem
\bibitem[Spott \latin{et~al.}(2017)Spott, Stanton, Volet, Peters, Meyer, and
  Bowers]{spott2017heterogeneous}
Spott,~A.; Stanton,~E.~J.; Volet,~N.; Peters,~J.~D.; Meyer,~J.~R.;
  Bowers,~J.~E. {Heterogeneous Integration for Mid-infrared Silicon Photonics}.
  \emph{{IEEE} Journal of Selected Topics in Quantum Electronics}
  \textbf{2017}, \emph{23}, 1--10\relax
\mciteBstWouldAddEndPuncttrue
\mciteSetBstMidEndSepPunct{\mcitedefaultmidpunct}
{\mcitedefaultendpunct}{\mcitedefaultseppunct}\relax
\EndOfBibitem
\bibitem[Wang \latin{et~al.}(2019)Wang, Haq, Sprengel, Malik, Vasiliev, Boehm,
  Simonyte, Vizbaras, Vizbaras, Campenhout, Baets, Amann, and
  Roelkens]{wang2019widely}
Wang,~R.; Haq,~B.; Sprengel,~S.; Malik,~A.; Vasiliev,~A.; Boehm,~G.;
  Simonyte,~I.; Vizbaras,~A.; Vizbaras,~K.; Campenhout,~J.~V.; Baets,~R.;
  Amann,~M.-C.; Roelkens,~G. Widely Tunable {III}{\textendash}V/Silicon Lasers
  for Spectroscopy in the Short-Wave Infrared. \emph{{IEEE} Journal of Selected
  Topics in Quantum Electronics} \textbf{2019}, \emph{25}, 1--12\relax
\mciteBstWouldAddEndPuncttrue
\mciteSetBstMidEndSepPunct{\mcitedefaultmidpunct}
{\mcitedefaultendpunct}{\mcitedefaultseppunct}\relax
\EndOfBibitem
\bibitem[Moutanabbir \latin{et~al.}(2021)Moutanabbir, Assali, Gong,
  O{\textquotesingle}Reilly, Broderick, Marzban, Witzens, Du, Yu, Chelnokov,
  Buca, and Nam]{moutanabbir2021monolithic}
Moutanabbir,~O.; Assali,~S.; Gong,~X.; O{\textquotesingle}Reilly,~E.;
  Broderick,~C.~A.; Marzban,~B.; Witzens,~J.; Du,~W.; Yu,~S.-Q.; Chelnokov,~A.;
  Buca,~D.; Nam,~D. Monolithic infrared silicon photonics: The rise of
  (Si){GeSn} semiconductors. \emph{Applied Physics Letters} \textbf{2021},
  \emph{118}, 110502\relax
\mciteBstWouldAddEndPuncttrue
\mciteSetBstMidEndSepPunct{\mcitedefaultmidpunct}
{\mcitedefaultendpunct}{\mcitedefaultseppunct}\relax
\EndOfBibitem
\bibitem[Reboud \latin{et~al.}(2024)Reboud, Concepción, Du, El~Kurdi,
  Hartmann, Ikonic, Assali, Pauc, Calvo, Cardoux, Kroemer, Coudurier,
  Rodriguez, Yu, Buca, and Chelnokov]{reboud2024advances}
Reboud,~V. \latin{et~al.}  Advances in GeSn alloys for MIR applications.
  \emph{Photonics and Nanostructures - Fundamentals and Applications}
  \textbf{2024}, \emph{58}, 101233\relax
\mciteBstWouldAddEndPuncttrue
\mciteSetBstMidEndSepPunct{\mcitedefaultmidpunct}
{\mcitedefaultendpunct}{\mcitedefaultseppunct}\relax
\EndOfBibitem
\bibitem[Cui \latin{et~al.}(2024)Cui, Zheng, Zhu, Liu, Wu, Huang, Yang, Liu,
  Liu, Zuo, and Cheng]{cui2024highspeed}
Cui,~J.; Zheng,~J.; Zhu,~Y.; Liu,~X.; Wu,~Y.; Huang,~Q.; Yang,~Y.; Liu,~Z.;
  Liu,~Z.; Zuo,~Y.; Cheng,~B. High-speed GeSn resonance cavity enhanced
  photodetectors for a 50 Gbps Si-based 2 $\mu$m band communication system.
  \emph{Photonics Research} \textbf{2024}, \emph{12}, 767\relax
\mciteBstWouldAddEndPuncttrue
\mciteSetBstMidEndSepPunct{\mcitedefaultmidpunct}
{\mcitedefaultendpunct}{\mcitedefaultseppunct}\relax
\EndOfBibitem
\bibitem[Atalla \latin{et~al.}(2024)Atalla, Lemieux-Leduc, Assali, Koelling,
  Daoust, and Moutanabbir]{atalla2024extended}
Atalla,~M. R.~M.; Lemieux-Leduc,~C.; Assali,~S.; Koelling,~S.; Daoust,~P.;
  Moutanabbir,~O. {Extended short-wave infrared high-speed all-GeSn PIN
  photodetectors on silicon}. \emph{APL Photonics} \textbf{2024},
  \emph{9}\relax
\mciteBstWouldAddEndPuncttrue
\mciteSetBstMidEndSepPunct{\mcitedefaultmidpunct}
{\mcitedefaultendpunct}{\mcitedefaultseppunct}\relax
\EndOfBibitem
\bibitem[Wang \latin{et~al.}(2022)Wang, Zhang, Zhang, Chen, Han, Huang, and
  Gong]{wang2022monolithic}
Wang,~H.; Zhang,~J.; Zhang,~G.; Chen,~Y.; Han,~K.; Huang,~Y.-C.; Gong,~X.
  {Monolithic Waveguide Group IV Multiple-Quantum-Well Photodetectors and
  Modulators on 300-mm Si Substrates for 2-$\mu$m Wavelength Optoelectronic
  Integrated Circuit}. \emph{IEEE Transactions on Electron Devices}
  \textbf{2022}, \emph{69}, 7161--7166\relax
\mciteBstWouldAddEndPuncttrue
\mciteSetBstMidEndSepPunct{\mcitedefaultmidpunct}
{\mcitedefaultendpunct}{\mcitedefaultseppunct}\relax
\EndOfBibitem
\bibitem[Li \latin{et~al.}(2020)Li, Wang, Liu, Chen, Du, Wang, Cai, Ma, and
  Yu]{li2020design}
Li,~X.~Y.; Wang,~J.~Y.; Liu,~Y.~F.; Chen,~J.~J.; Du,~Y.; Wang,~W.; Cai,~Y.;
  Ma,~J.~P.; Yu,~M.~B. Design of Ge$_{1-x}$Sn$_x$-on-Si waveguide
  photodetectors featuring high-speed high-sensitivity photodetection in the C-
  to U-bands. \emph{Applied Optics} \textbf{2020}, \emph{59}, 7646\relax
\mciteBstWouldAddEndPuncttrue
\mciteSetBstMidEndSepPunct{\mcitedefaultmidpunct}
{\mcitedefaultendpunct}{\mcitedefaultseppunct}\relax
\EndOfBibitem
\bibitem[Ghosh \latin{et~al.}(2022)Ghosh, Bansal, Sun, Soref, Cheng, and
  Chang]{ghosh2022design}
Ghosh,~S.; Bansal,~R.; Sun,~G.; Soref,~R.~A.; Cheng,~H.-H.; Chang,~G.-E. Design
  and Optimization of GeSn Waveguide Photodetectors for 2-$\mu$m Band Silicon
  Photonics. \emph{Sensors} \textbf{2022}, \emph{22}, 3978\relax
\mciteBstWouldAddEndPuncttrue
\mciteSetBstMidEndSepPunct{\mcitedefaultmidpunct}
{\mcitedefaultendpunct}{\mcitedefaultseppunct}\relax
\EndOfBibitem
\bibitem[Tsai \latin{et~al.}(2021)Tsai, Lin, Cheng, Lee, Cheng, and
  Chang]{tsai2021gesn}
Tsai,~C.-H.; Lin,~K.-C.; Cheng,~C.-Y.; Lee,~K.-C.; Cheng,~H.~H.; Chang,~G.-E.
  {GeSn} lateral p-i-n waveguide photodetectors for mid-infrared integrated
  photonics. \emph{Optics Letters} \textbf{2021}, \emph{46}, 864\relax
\mciteBstWouldAddEndPuncttrue
\mciteSetBstMidEndSepPunct{\mcitedefaultmidpunct}
{\mcitedefaultendpunct}{\mcitedefaultseppunct}\relax
\EndOfBibitem
\bibitem[Wang \latin{et~al.}(2021)Wang, Zhang, Zhang, Chen, Huang, and
  Gong]{wang2021highspeed}
Wang,~H.; Zhang,~J.; Zhang,~G.; Chen,~Y.; Huang,~Y.-C.; Gong,~X. High-speed and
  high-responsivity p-i-n waveguide photodetector at a 2 $\mu$m wavelength with
  a Ge$_{0.92}$Sn$_{0.08}$/Ge multiple-quantum-well active layer. \emph{Optics
  Letters} \textbf{2021}, \emph{46}, 2099\relax
\mciteBstWouldAddEndPuncttrue
\mciteSetBstMidEndSepPunct{\mcitedefaultmidpunct}
{\mcitedefaultendpunct}{\mcitedefaultseppunct}\relax
\EndOfBibitem
\bibitem[Xu \latin{et~al.}(2019)Xu, Han, Huang, Lee, Kang, Masudy-Panah, Wu,
  Lei, Zhao, Wang, Tan, Gong, and Yeo]{xu2019integrating}
Xu,~S.; Han,~K.; Huang,~Y.-C.; Lee,~K.~H.; Kang,~Y.; Masudy-Panah,~S.; Wu,~Y.;
  Lei,~D.; Zhao,~Y.; Wang,~H.; Tan,~C.~S.; Gong,~X.; Yeo,~Y.-C. Integrating
  {GeSn} photodiode on a 200 mm Ge-on-insulator photonics platform with Ge
  {CMOS} devices for advanced {OEIC} operating at 2 µm band. \emph{Optics
  Express} \textbf{2019}, \emph{27}, 26924\relax
\mciteBstWouldAddEndPuncttrue
\mciteSetBstMidEndSepPunct{\mcitedefaultmidpunct}
{\mcitedefaultendpunct}{\mcitedefaultseppunct}\relax
\EndOfBibitem
\bibitem[Joo \latin{et~al.}(2021)Joo, Kim, Burt, Jung, Zhang, Chen, Parluhutan,
  Kang, Lee, Assali, Ikonic, Moutanabbir, Cho, Tan, and Nam]{Joo2021}
Joo,~H.-J.; Kim,~Y.; Burt,~D.; Jung,~Y.; Zhang,~L.; Chen,~M.;
  Parluhutan,~S.~J.; Kang,~D.-H.; Lee,~C.; Assali,~S.; Ikonic,~Z.;
  Moutanabbir,~O.; Cho,~Y.-H.; Tan,~C.~S.; Nam,~D. 1D photonic crystal direct
  bandgap GeSn-on-insulator laser. \emph{Applied Physics Letters}
  \textbf{2021}, \emph{119}, 201101\relax
\mciteBstWouldAddEndPuncttrue
\mciteSetBstMidEndSepPunct{\mcitedefaultmidpunct}
{\mcitedefaultendpunct}{\mcitedefaultseppunct}\relax
\EndOfBibitem
\bibitem[Jung \latin{et~al.}(2022)Jung, Burt, Zhang, Kim, Joo, Chen, Assali,
  Moutanabbir, Tan, and Nam]{Jung22}
Jung,~Y.; Burt,~D.; Zhang,~L.; Kim,~Y.; Joo,~H.-J.; Chen,~M.; Assali,~S.;
  Moutanabbir,~O.; Tan,~C.~S.; Nam,~D. Optically pumped low-threshold microdisk
  lasers on a GeSn-on-insulator substrate with reduced defect density.
  \emph{Photon. Res.} \textbf{2022}, \emph{10}, 1332--1337\relax
\mciteBstWouldAddEndPuncttrue
\mciteSetBstMidEndSepPunct{\mcitedefaultmidpunct}
{\mcitedefaultendpunct}{\mcitedefaultseppunct}\relax
\EndOfBibitem
\bibitem[Atalla \latin{et~al.}(2021)Atalla, Assali, Attiaoui, Lemieux‐Leduc,
  Kumar, Abdi, and Moutanabbir]{atalla2021allgroup}
Atalla,~M. R.~M.; Assali,~S.; Attiaoui,~A.; Lemieux‐Leduc,~C.; Kumar,~A.;
  Abdi,~S.; Moutanabbir,~O. {All-Group IV Transferable Membrane Mid-Infrared
  Photodetectors}. \emph{Advanced Functional Materials} \textbf{2021},
  \emph{31}, 2006329\relax
\mciteBstWouldAddEndPuncttrue
\mciteSetBstMidEndSepPunct{\mcitedefaultmidpunct}
{\mcitedefaultendpunct}{\mcitedefaultseppunct}\relax
\EndOfBibitem
\bibitem[Kim \latin{et~al.}(2022)Kim, Assali, Burt, Jung, Joo, Chen, Ikonic,
  Moutanabbir, and Nam]{kim2022}
Kim,~Y.; Assali,~S.; Burt,~D.; Jung,~Y.; Joo,~H.-J.; Chen,~M.; Ikonic,~Z.;
  Moutanabbir,~O.; Nam,~D. Enhanced GeSn Microdisk Lasers Directly Released on
  Si. \emph{Advanced Optical Materials} \textbf{2022}, \emph{10}, 2101213\relax
\mciteBstWouldAddEndPuncttrue
\mciteSetBstMidEndSepPunct{\mcitedefaultmidpunct}
{\mcitedefaultendpunct}{\mcitedefaultseppunct}\relax
\EndOfBibitem
\bibitem[Chen \latin{et~al.}(2022)Chen, Wu, Zhang, Zhou, Fan, and
  Tan]{chen2022transferable}
Chen,~Q.; Wu,~S.; Zhang,~L.; Zhou,~H.; Fan,~W.; Tan,~C.~S. Transferable
  single-layer GeSn nanomembrane resonant-cavity-enhanced photodetectors for 2
  ${\mu}$m band optical communication and multi-spectral short-wave infrared
  sensing. \emph{Nanoscale} \textbf{2022}, \emph{14}, 7341--7349\relax
\mciteBstWouldAddEndPuncttrue
\mciteSetBstMidEndSepPunct{\mcitedefaultmidpunct}
{\mcitedefaultendpunct}{\mcitedefaultseppunct}\relax
\EndOfBibitem
\bibitem[Lemieux-Leduc \latin{et~al.}(2025)Lemieux-Leduc, Atalla, Assali,
  Koelling, Daoust, Luo, Daligou, Brodeur, Kéna-Cohen, Peter, and
  Moutanabbir]{lemieuxleduc2025transferprinted}
Lemieux-Leduc,~C.; Atalla,~M. R.~M.; Assali,~S.; Koelling,~S.; Daoust,~P.;
  Luo,~L.; Daligou,~G.; Brodeur,~J.; Kéna-Cohen,~S.; Peter,~Y.-A.;
  Moutanabbir,~O. {Transfer-Printed Multiple GeSn Membrane Mid-Infrared
  Photodetectors}. \emph{IEEE Journal of Selected Topics in Quantum
  Electronics} \textbf{2025}, \emph{31}, 1--10\relax
\mciteBstWouldAddEndPuncttrue
\mciteSetBstMidEndSepPunct{\mcitedefaultmidpunct}
{\mcitedefaultendpunct}{\mcitedefaultseppunct}\relax
\EndOfBibitem
\bibitem[Carlson \latin{et~al.}(2012)Carlson, Bowen, Huang, Nuzzo, and
  Rogers]{carlson2012transfer}
Carlson,~A.; Bowen,~A.~M.; Huang,~Y.; Nuzzo,~R.~G.; Rogers,~J.~A. Transfer
  Printing Techniques for Materials Assembly and Micro/Nanodevice Fabrication.
  \emph{Advanced Materials} \textbf{2012}, \emph{24}, 5284--5318\relax
\mciteBstWouldAddEndPuncttrue
\mciteSetBstMidEndSepPunct{\mcitedefaultmidpunct}
{\mcitedefaultendpunct}{\mcitedefaultseppunct}\relax
\EndOfBibitem
\bibitem[Yoon \latin{et~al.}(2015)Yoon, Lee, Kang, Meitl, Bower, and
  Rogers]{yoon2015heterogeneously}
Yoon,~J.; Lee,~S.; Kang,~D.; Meitl,~M.~A.; Bower,~C.~A.; Rogers,~J.~A.
  Heterogeneously Integrated Optoelectronic Devices Enabled by Micro‐Transfer
  Printing. \emph{Advanced Optical Materials} \textbf{2015}, \emph{3},
  1313--1335\relax
\mciteBstWouldAddEndPuncttrue
\mciteSetBstMidEndSepPunct{\mcitedefaultmidpunct}
{\mcitedefaultendpunct}{\mcitedefaultseppunct}\relax
\EndOfBibitem
\bibitem[De~Groote \latin{et~al.}(2016)De~Groote, Cardile, Subramanian,
  Fecioru, Bower, Delbeke, Baets, and Roelkens]{degroote2016transferprinting}
De~Groote,~A.; Cardile,~P.; Subramanian,~A.~Z.; Fecioru,~A.~M.; Bower,~C.;
  Delbeke,~D.; Baets,~R.; Roelkens,~G. Transfer-printing-based integration of
  single-mode waveguide-coupled III-V-on-silicon broadband light emitters.
  \emph{Optics Express} \textbf{2016}, \emph{24}, 13754\relax
\mciteBstWouldAddEndPuncttrue
\mciteSetBstMidEndSepPunct{\mcitedefaultmidpunct}
{\mcitedefaultendpunct}{\mcitedefaultseppunct}\relax
\EndOfBibitem
\bibitem[Zhang \latin{et~al.}(2019)Zhang, Muliuk, Juvert, Kumari, Goyvaerts,
  Haq, Op~de Beeck, Kuyken, Morthier, Van~Thourhout, Baets, Lepage, Verheyen,
  Van~Campenhout, Gocalinska, O’Callaghan, Pelucchi, Thomas, Corbett,
  Trindade, and Roelkens]{zhang2019iiivonsi}
Zhang,~J. \latin{et~al.}  III-V-on-Si photonic integrated circuits realized
  using micro-transfer-printing. \emph{APL Photonics} \textbf{2019}, \emph{4},
  110803\relax
\mciteBstWouldAddEndPuncttrue
\mciteSetBstMidEndSepPunct{\mcitedefaultmidpunct}
{\mcitedefaultendpunct}{\mcitedefaultseppunct}\relax
\EndOfBibitem
\bibitem[Haq \latin{et~al.}(2020)Haq, Vaskasi, Zhang, Gocalinska, Pelucchi,
  Corbett, and Roelkens]{haq2020microtransferprinted}
Haq,~B.; Vaskasi,~J.~R.; Zhang,~J.; Gocalinska,~A.; Pelucchi,~E.; Corbett,~B.;
  Roelkens,~G. Micro-transfer-printed {III}-V-on-silicon C-band distributed
  feedback lasers. \emph{Optics Express} \textbf{2020}, \emph{28}, 32793\relax
\mciteBstWouldAddEndPuncttrue
\mciteSetBstMidEndSepPunct{\mcitedefaultmidpunct}
{\mcitedefaultendpunct}{\mcitedefaultseppunct}\relax
\EndOfBibitem
\bibitem[Mao \latin{et~al.}(2004)Mao, Li, Zuo, Cheng, Teng, Luo, Yu, and
  Wang]{mao2004fabrication}
Mao,~R.; Li,~C.; Zuo,~Y.; Cheng,~B.; Teng,~X.; Luo,~L.; Yu,~J.; Wang,~Q.
  {Fabrication of 1.55-$\mu$m Si-Based Resonant Cavity Enhanced Photodetectors
  Using Sol-Gel Bonding}. \emph{IEEE Photonics Technology Letters}
  \textbf{2004}, \emph{16}, 1930--1932\relax
\mciteBstWouldAddEndPuncttrue
\mciteSetBstMidEndSepPunct{\mcitedefaultmidpunct}
{\mcitedefaultendpunct}{\mcitedefaultseppunct}\relax
\EndOfBibitem
\bibitem[Atalla \latin{et~al.}(2022)Atalla, Assali, Koelling, Attiaoui, and
  Moutanabbir]{atalla2022highbandwidth}
Atalla,~M. R.~M.; Assali,~S.; Koelling,~S.; Attiaoui,~A.; Moutanabbir,~O.
  {High-Bandwidth Extended-SWIR GeSn Photodetectors on Silicon Achieving
  Ultrafast Broadband Spectroscopic Response}. \emph{ACS Photonics}
  \textbf{2022}, \emph{9}, 1425--1433\relax
\mciteBstWouldAddEndPuncttrue
\mciteSetBstMidEndSepPunct{\mcitedefaultmidpunct}
{\mcitedefaultendpunct}{\mcitedefaultseppunct}\relax
\EndOfBibitem
\bibitem[Hattasan \latin{et~al.}(2011)Hattasan, Gassenq, Cerutti, Rodriguez,
  Tournie, and Roelkens]{hattasan2011heterogeneous}
Hattasan,~N.; Gassenq,~A.; Cerutti,~L.; Rodriguez,~J.-B.; Tournie,~E.;
  Roelkens,~G. Heterogeneous Integration of GaInAsSb p-i-n Photodiodes on a
  Silicon-on-Insulator Waveguide Circuit. \emph{IEEE Photonics Technology
  Letters} \textbf{2011}, \emph{23}, 1760--1762\relax
\mciteBstWouldAddEndPuncttrue
\mciteSetBstMidEndSepPunct{\mcitedefaultmidpunct}
{\mcitedefaultendpunct}{\mcitedefaultseppunct}\relax
\EndOfBibitem
\end{mcitethebibliography}

\newpage

\setcounter{page}{1}

\section*{Supporting Information: Waveguide-Coupled Mid-Infrared GeSn Membrane Photodetectors on Silicon-on-Insulator}

Cédric~Lemieux-Leduc, Mahmoud~R.~M.~Atalla, Simone~Assali, Nicolas~Rotaru, Julien~Brodeur, Stéphane~Kéna-Cohen, Oussama~Moutanabbir*, and Yves-Alain~Peter
\\
\textit{Department of Engineering Physics, Polytechnique Montréal, Montréal, Canada}
\\
\textit{* E-mail: oussama.moutanabbir@polymtl.ca}
\\
The Supporting Information contains 4 pages excluding the cover page. It includes 4 figures.

\newpage

\subsection*{Extraction of optical constants using ellipsometry}

The optical constants of the Ge$_{0.89}$Sn$_{0.11}$ material (Fig. S1) and SU-8 (Fig. S2) used in the modeling were extracted via spectroscopic ellipsometry using an RCA2 ellipsometer from J.A. Woollam, which provides data from 0.4 to 2.5~$\mu$m. This technique measures changes in the polarization state of an elliptically polarized light source upon reflection from the thin film, allowing interactions with different layers to be analyzed. For Ge$_{0.89}$Sn$_{0.11}$, a sum of oscillators was employed to model and extract the optical constants, while a Cauchy function was used for SU-8. Only the refractive index is only shown for SU-8, as its extinction coefficient was measured to be negligible from the visible up to 2.5~$\mu$m.

Since ellipsometry could not be performed on small-sized membranes, the optical constants from the as-grown sample were used as an approximation. While strain relaxation in membranes alters the alloy's band structure, enhancing absorption by extending the absorption cutoff, using the as-grown sample is expected to provide similar refractive index values. Based on previous observations in [1], the absorption cutoff is estimated to shift from approximately 2.8~$\mu$m to 3.1~$\mu$m in the membrane.

\renewcommand{\thefigure}{S1}
\begin{figure}[htbp]
    \centering
    \includegraphics[width=\textwidth]{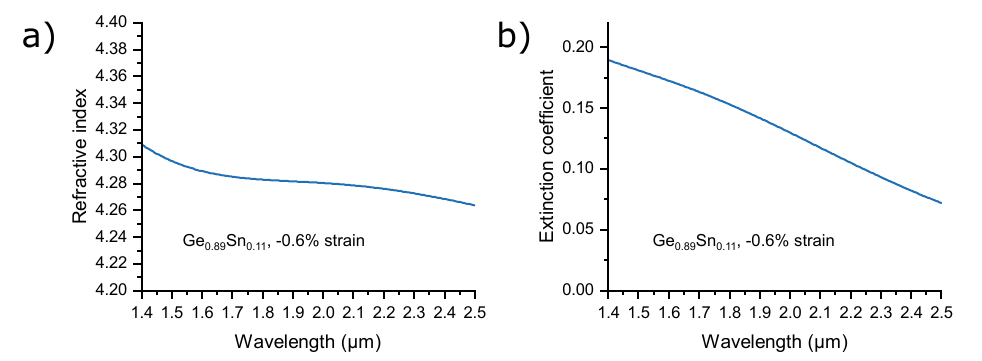}
    \caption{Optical constants of Ge$_{0.89}$Sn$_{0.11}$ at a compressive strain of -0.6\% extracted using ellipsometry. (a) Refractive index. (b) Extinction coefficient.}
    \label{fig:nk_gesn}
\end{figure}

\renewcommand{\thefigure}{S2}
\begin{figure}[htbp]
    \centering
    \includegraphics[width=\textwidth]{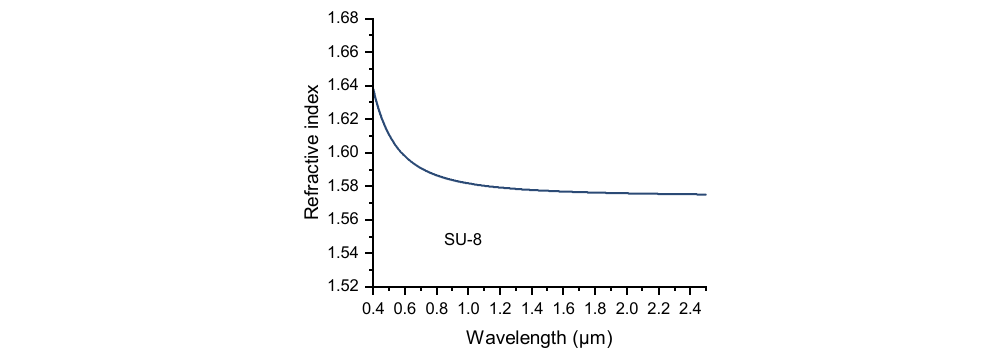}
    \caption{Refractive index of SU-8 extracted using ellipsometry. The extinction coefficient is negligible for the wavelength range of interest.}
    \label{fig:nk_su8}
\end{figure}

\subsection*{Comparison between evaporated and transfer-printed metal contacts}

The Ohmic-like behavior observed in Fig. 5(a) contrasts with the membrane devices fabricated in [1], where contacts were exfoliated and transfer-printed rather than evaporated. Figure S3 compares the I-V curves obtained using these two techniques, showing that the evaporated contacts exhibit more linear characteristics and enhanced conductivity. The difference in deposition methods suggests that metal evaporation may have partially fused the metal with the semiconductor, preventing the formation of a Schottky barrier.

\renewcommand{\thefigure}{S3}
\begin{figure}[htbp]
    \centering
    \includegraphics[width=0.7\textwidth]{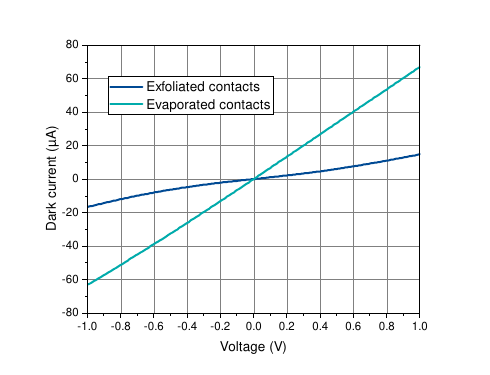}
    \caption{Current-voltage (I-V) measurements of a device with evaporated contacts compared to exfoliated contacts.}
    \label{fig:exf_v_eva_contacts}
\end{figure}

\subsection*{Modeling of the generation rate}

To evaluate the efficiency of the fabricated devices, the photocarrier generation rate was computed from FDTD simulations by recording the electric field $E$ within the absorbing medium (GeSn) with the complex dielectric constant $\epsilon$. The generation rate $G$ is defined as:

\newcommand{\IM}{\mathrm{Im}}
\begin{equation}
    G = -\omega|E|^2\IM(\epsilon)
\end{equation}

where $\omega$ is the angular frequency. Figure S4 presents the 2D and cumulative generation rates for designs EC and ECGAC along the length of the 20~$\mu$m membrane, with a separation gap of 180 nm and SU-8 as the surrounding medium. Simulations were performed using Lumerical FDTD with a mesh accuracy setting of 6, the mesh type set to auto non-uniform, and a minimum mesh size of 50 nm.

For Design ECGAC, the first half of the membrane overlaps the tapered waveguide, while the second half overlaps the diffraction grating. Although the waveguide is tapered in Design ECGAC, it does not significantly affect the 2D rate compared to Design EC in the first 10~$\mu$m. The cumulative generation rates correspond to the absorbed power fraction after normalization to the incident power. The final cumulative rate values (0.85 for Design EC and 0.88 for Design ECGAC) are similar after light has interacted with the entire membrane, suggesting that evanescent coupling is likely the dominant coupling mechanism.

\renewcommand{\thefigure}{S4}
\begin{figure}[htbp]
    \centering
    \includegraphics[width=0.7\textwidth]{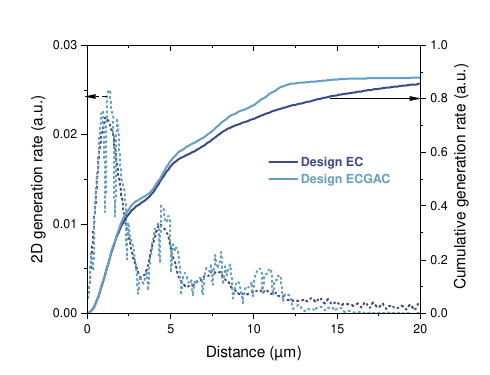}
    \caption{Comparison of the computed 2D and cumulative generation rates for designs EC and ECGAC.}
    \label{fig:G_rate_comparison}
\end{figure}

\section*{References}

\begin{itemize}
    \item[{[1]}] Lemieux-Leduc, C.; Atalla, M. R. M; Assali, S.; Koelling, S.; Daoust, P.; Luo, L.; Daligou, G.; Brodeur, J.; Kéna-Cohen, S.; Peter, Y.-A.; Moutanabbir, O. Transfer-Printed Multiple GeSn Membrane Mid-Infrared Photodetectors. \textit{IEEE Journal of Selected Topics in Quantum Electronics} \textbf{2025}, 31, 1-10.
\end{itemize}

\end{document}